# Reliability Considerations for the Operation of Large Accelerator User Facilities

*F. J. Willeke*
Brookhaven National Laboratory, Upton, NY

**Abstract**
The lecture provides an overview of considerations relevant for achieving highly reliable operation of accelerator based user facilities. The article starts with an overview of statistical reliability formalism which is followed by high reliability design considerations with examples. The article closes with operational aspects of high reliability such as preventive maintenance and spares inventory.

## 1  Introduction

In previous decades, accelerators were developed and optimized as tools to explore the energy frontier for studying sub nuclear particles. However, more recently, another aspect of accelerator optimization has become more important, which is highly reliable operations to produce a large quantity of particle collisions ('particle factories') or photons (light sources), serving a large and diverse user community. The reliability aspect is particularly relevant for light sources. Light sources have large user communities of several thousand users organized in small independent research teams, each of which uses only a small fraction of the beam time. Even small operational inefficiencies due to frequent failures and interruptions might cause the total loss of allocated beam time of some research teams, with significant disruption of their science programmes. For these reasons, an increasing emphasis has been put on highly reliable operations. Reliability is usually defined as the total relative amount of beam time made available to users within the scheduled time period. A reliability of 95% is considered a tolerable lower limit for modern light sources. Reliability values of the order of 98% are reported frequently and are not an unusual achievement. This means that for a scheduled yearly beam time of for example 5000 h, only 250 h or less of user operations may be lost due to failures. Assuming that, on average, full recovery from a failure requires two hours, the time between interruptions must be larger than 40 h (assuming 24 h/d and 7 d/week of operation) on average. Science with synchrotron radiation has become very sophisticated and the delivery of a beam is not a sufficient criterion for reliability any more. Users need a beam of the planned beam energy and with nearly constant intensity, high spatial stability and high reproducibility of all beam parameters after changes of operational mode, such as changes in photon energy by changing the field strengths of undulator magnets. Accelerators consist of a large number of active components, many of them with high power consumption, which must function simultaneously to enable beam operation. They are connected and coordinated by sophisticated digital controls, and precision timing is usually a condition for proper functioning. For a facility with 100,000 of such components, any component may fail only after $4 \times 10^6$ h of operation.

In the past, an accelerator facility needed several years or even a decade of operations to mature operations and develop the hardware system before such demanding operating goals could be realized. What appears to be desirable is to develop requirements to be taken into account in the design of the accelerator and in planning for its operations. Thus, accelerators must be designed for high reliability. When operating the facilities, all components must be maintained carefully, based on a comprehensive preventive maintenance programme to minimize unscheduled downtime. This involves the monitoring of all components over time to identify any deviation from normal functioning, so as to prevent a failure during operations by timely repair or replacement.

These topics are discussed in this paper, which is organized as follows:
- introduction to reliability theory and definition of relevant parameters and properties;
- aspects and examples for high-reliability design;
- maintenance programmes;
- spares management.

## 2   Short summary of reliability definitions and relationships

### 2.1   General remarks

The purpose of this section is to show how reliability relevant parameters and functions that are used to analyse failure statistics and to make reliability predictions are related to observable and measurable quantities. To provide some understanding of the underlying statistical nature of the relationships, a short derivation of the most important formulae from basic principles and assumptions is presented.

In many areas of physics and engineering, the behaviour of complex, though deterministic, systems is described successfully by a statistical model in which events are considered random and independent, as insufficient detailed information about these systems is available. Random events are considered independent. This implies that the probability for a failure is independent of the history of previous failures or the failure of other components. While this is a rather practical and successful approach in most cases, we must keep in mind that we are dealing with deterministic systems and we are using the concept of statistics as a model. In particular, it should be noted at this point that the number of components in the systems considered is small, so the uncertainty of statistical prediction is considerable. Moreover, one needs to be aware that failures and different times or in different subsystems are not independent, adding another element of uncertainty to the outcome of statistical modelling.

Let us consider the following example:

A circuit breaker trips due to external overvoltage and a large number of magnet power supplies lose power and trip. During recovery from the event, one supply is found to be damaged from that occurrence and needs to be repaired. The two failures, the circuit breaker trip and the power supply failure, are clearly not independent. The power supply would most likely not have failed and continued to function for a while. However, it is also quite likely that the power supply would have failed at a somewhat later time during normal operating procedures, such as turn-off–turn-on cycles because there must have been a hidden defect, as power supplies are designed to survive power failures. Such events might be considered quasi-independent. However, there are strongly dependent failures, such as a large cooling-water leak causing water to penetrate a power supply, which may lead to corrosion and subsequent failure. Such events are usually fairly rare and do not have a large statistical significance. For now, we will assume that there are no significant dependencies of failures and that the statistical failure model is adequate.

### 2.2   Mean time between failure

The mean time between failures (MTBF) is an important observable parameter, which can be related to other statistical functions and parameters related to failure occurrence and system reliability. For a single component, it is simply defined as the average time between two failure events, which is the number of failures in a certain time period. This assumes that the system can be restored or repaired after a failure event. For non-repairable components one just averages the time to failure for a number of components. This is quite intuitive and, as we will see, the equivalency of these two definitions can be shown. To assess the impact of failure, another quantity is quite relevant; the time it takes on average to return a failed system to service, the mean time to repair (MTTR). The availability of a repairable system is defined as

$$\text{Availability} = 1 - \frac{\text{MTTR}}{\text{MTBF} + \text{MTTR}}.$$

Where a system consists of different constituents or subsystems (labelled here by index *i*), the availability can be written as

$$\text{Availability} = \prod_i \left[1 - \frac{\text{MTTR}_i}{\text{MTBF}_i + \text{MTTR}_i}\right] \cong 1 - \sum_i \frac{\text{MTTR}_i}{\text{MTBF}_i + \text{MTTR}_i}, \text{for MTBF} \gg \text{MTTR}.$$

To predict statistical failure behaviour, it is useful to introduce the concept of an instantaneous failure probability, which is closely related to the MTBF. We define *p* as the probability for a failure to occur in a small interval of time $\Delta t$. For the time being, we assume that *p* is the same for any interval of time,

$$p = \lambda \cdot \Delta t,$$

where $\lambda$ is called the instantaneous failure rate. If $\lambda$ is constant in time, the failure density distribution, *f*, or the probability of surviving a certain number $n - 1$ of time intervals but failing in the *n*th time interval is

$$f_n \Delta t = [1 - p]^{n-1} p,$$

and $f_n$ is a normalized distribution with

$$\sum_{n=1}^{\infty} f_n = 1,$$

using

$$\sum_{n=0}^{\infty} q^n = \frac{1}{1-q} \text{ for } |q| < 1.$$

The MTBF can be interpreted as the expectation value of the time to failure for the density distribution, thus

$$MTTBF = <n> \Delta t = \sum_{n=1}^{\infty} n \cdot [1-p]^{n-1} p \cdot \Delta t =$$
$$-p \cdot \Delta t \cdot \frac{d}{dp} \sum_{n=1}^{\infty} [1-p]^{n-1} p \cdot \Delta t = -p \cdot \Delta t \frac{d}{dp} \frac{1}{p} = \frac{\Delta t}{p} = \frac{1}{\lambda}.$$

This is an important relationship, which allows us to relate the parameters of statistical functions to observations: MTBF = $1/\lambda$. Note that this simple relationship holds only for time-independent instantaneous failure rates but that an equivalent relationship can be given for more complicated cases of time dependent $\lambda(t)$ (called the hazard function) as well.

## 2.3 Failure and survival functions

The failure and survival functions are important tools to predict failures. The failure function $F_N$ gives the probability that failure occurs within a time $N \cdot \Delta t$. It is given as the sum over the failure distribution density up to a time $N \cdot \Delta t$:

$$F_N = \sum_{n=1}^{N} f_n = 1 - (1-p)^N .$$

The survival function $S_N$ is the complement of the failure function and gives the probability that the system will survive for a time $N \cdot \Delta t$.

$$S_N = 1 - F_N .$$

## 2.4 Systems with identical components

Accelerators are built with subsystems which contain identical components. The considerations of the previous section can be generalized to include multicomponent systems. Assume a system of $N$ identical components, each component having an instantaneous failure probability of $\lambda \cdot \Delta t$; ($\lambda$ is also called the hazard function). The probability of $m$ components failing during a time $\Delta t$ is then:

$$P_{Nm} = \binom{N}{m} \cdot (1-p)^{N-m} p^m .$$

Note that $P_{Nm}$ is a normalized distribution function with

$$\sum_{m=1}^{N} P_{Nm} = (q+p)^N |_{q \to 1-p} = 1 .$$

The average number of failed components within a time interval $\Delta t$ is, as might have been expected:

$$\langle m \rangle = \sum_{m=1}^{N} P_{Nm} \cdot m = p \frac{d}{dp}(q+p)^N |_{q \to 1-p} = N \cdot p .$$

The likelihood to have no failure in the time interval is $P_{N0} = (1-p)^N$. When computing the MTBF for the system with $N$ components, $1-p$ in the equation for the MTBF for one component in Section 2.2 has to be replaced by $(1-p)^N$:

$$\text{MTBF} = \frac{\Delta t}{1-(1-p)^N} \cong \frac{\Delta t}{N \cdot p} = \frac{1}{N \cdot \lambda} .$$

Thus the MTBF for the system with $N$ components is $N$ times smaller than for a single-component system as one might have assumed intuitively.

## 2.5 Non-constant failure rates

So far, we have assumed that the probability for failure within $\Delta t$ is constant in time. However, there are many reasons for a non-constant instantaneous failure rate. New systems have a certain fraction of components with hidden defects, leading to enhanced failure rates early in the life cycle. Many components wear-out or age as a result of other effects (damage due to repetitive high temperature, accumulation of dust and aggressive chemicals, change of material properties, such as elasticity, with time and so on). Another reason for time dependence of failures is changing external conditions, such as temperature and humidity. Systems with high voltage, for example, tend to develop arcs if the humidity of the air changes. Another parameter is the time since the last maintenance, during which mechanical clamps might loosen or dust might have accumulated.

For these reasons, to describe real systems, we must develop the formalism under the assumption that $\lambda$ is not constant but may depend on time.

We start with the assumption that $p$ is the probability for a failure within a short interval of time $\Delta t$, but that the probability may vary for different time intervals labelled $n$. Thus $p \to p_n = \lambda_n \cdot \Delta t$. The failure density distribution then becomes

$$f_N \cdot \Delta t = \lambda_N \cdot \Delta t \cdot \prod_{n=1}^{N}(1 - \lambda_n \cdot \Delta t).$$

The expression for *f* is more conveniently represented by a continuous function by making the time steps infinitesimally small. To make this transition we rewrite the probability density distribution *f* as:

$$f_N = \lambda_N \cdot \exp\left[\sum_{n=1}^{N} \ln(1 - \lambda_n \Delta t)\right].$$

At this point, we can make the transition $\Delta t \to 0$ and write, correspondingly,

$$f_N = \lim_{N \to \infty}\left\{\lambda_N \cdot \exp\left[\sum_{n=1}^{N} \ln(1 - \lambda_n \Delta t)\right]\right\}.$$

Since $\lambda_n \Delta t$ is approaching zero, the logarithm can be expressed in terms of its Taylor expansion,

$$\ln(1 - \lambda_n \Delta t) \to -\lambda_n \Delta t,$$

resulting in

$$f_N = \lim_{N \to 0}\left\{\lambda_N \cdot \exp\left[\sum_{n=1}^{N} -\lambda_n \Delta t\right]\right\},$$

and this leads to

$$f(t) = \lambda(t) \cdot \exp\left[-\int_0^t \lambda(\tau) d\tau\right].$$

The failure function *F(t)*, which gives the probability of failure within the interval [0,*t*] is the integral over the probability density distribution *f(t)*:

$$F(t) = \int_0^t d\theta \lambda(\theta) \cdot \exp\left[-\int_0^\theta \lambda(\tau) d\tau\right] = 1 - \exp\left[-\int_0^t \lambda(\tau) d\tau\right].$$

The complement of the failure function is the survival function *S(t)*, which gives the probability of surviving a time *t* without failure:

$$S(t) = \exp\left[-\int_0^t \lambda(\tau) d\tau\right].$$

The instantaneous failure rate may be expressed by *F(t)* and *S(t)*:

$$\lambda(t) = \frac{\frac{d}{dt}F(t)}{S(t)}.$$

Now we can express the MTBF in terms of the continuous functions that we have derived. The MTBF is the expectation value of the time until failure using the probability density distribution,

$$\text{MTBF} = \int_0^\infty dt\, t \cdot f(t)$$

$$= \int_0^\infty dt\, t \cdot \lambda(t) \cdot \exp\left[-\int_0^t \lambda(\tau)\, d\tau\right]$$

$$= \left[t \cdot \exp\left[-\int_0^t \lambda(\tau)\, d\tau\right]\right]_0^\infty + \int_0^\infty dt \exp\left[-\int_0^t \lambda(\tau)\, d\tau\right]$$

$$= \int_0^\infty dt\, S(t).$$

For example, if $\lambda$ constant:

$$\text{MFBF} = \int_0^\infty dt \exp\left[-\int_0^t \lambda\, d\tau\right] = \int_0^\infty dt \exp[-\lambda t] = \frac{1}{\lambda}.$$

With these functions, we are now able to calculate an expectation value for the medium residual lifetime MRL(*t*) of a system that has survived a certain time *t* without failure. We use a similar expression to that for calculating the MTBF, except that the integral now extends from time *t* to infinity and the expression is divided by the probability of survival up to the time *t*, since only cases that have survived up to time = *t* are being considered

$$\text{MRL}(t) = \frac{\int_0^\infty f(t+\tau) \cdot \tau \cdot d\tau}{S(t)},$$

$$\text{MRL}(t) = \frac{\int_0^\infty d\tau\, S(t+\tau)}{S(t)}.$$

Note that for statistical failures, i.e., $S(t) = \exp(-\lambda t)$, MRL = $1/\lambda$, which is identical to MTBF.

## 2.6 Statistical modelling of real systems

A useful parameterization for describing the failure statistics of real systems is the Weibull parameterization. The parameters of the Weibull model can be chosen so as to describe any of the three failure modes discussed so far: premature failure, statistical failure, and wear-out or ageing-related failure. Weibull was a Swedish engineer who introduced his model in the 1930s [1] in the context of describing fatigue and wear-out and this model has been applied to many use-cases. Since then, many modified or alternative parameterizations have been proposed and successfully applied (see, for example, Ref. [2], and quotations therein). However, it would stray too far from the purpose of this lecture to discuss them all. In the Weibull model, the instantaneous failure probability function $\lambda(t)$ is described as

$$\lambda(t) = \frac{a}{b}\left(\frac{t}{b}\right)^{a-1}$$

The probability density distribution for failures in the Weibull model is

$$f(t) = \frac{a}{b}\left(\frac{t}{b}\right)^{a-1} \cdot \exp\left[-\left(\frac{t}{b}\right)^a\right].$$

The probability for failure is then

$$F(t) = 1 - \exp\left[-\left(\frac{t}{b}\right)^a\right].$$

The parameter $b$ is a lifetime parameter, which for $a = 1$ equals the MTBF. The parameter $a$ describes the nature of the failure statistics. If $a < 1$, the model describes early, premature, failure, where the failure probability decreases with increasing time; $a = 0$ describes the case where the failure probability function is constant, which is referred to as the statistical failure mode; $a > 1$ describes an increasing failure rate with increasing time, as expected for wear-out and ageing. If $a > 1$, the stronger the deviation from $a = 1$, the sharper the probability density distribution $f(t)$ peaks around the lifetime value $t = b$. For $a < 1$, the closer $a$ is to zero, the faster the decay but the slower the approach of probability to failure to $F = 1$. A real system has elements of all three phases of failure. The hazard function for various values of $a$ is depicted in Fig. 1.

Given an inventory of identical components, a certain fraction, $c_1$, will fail prematurely, another fraction, $c_2$, will fail statistically and the majority of the components, $c_3$, will fail due to wear-out and ageing. The total probability for failure is a weighted sum of the three components,

$$F(t) = \sum_{i=1}^{3} c_i \cdot \left\{ 1 - \exp\left[-\left(\frac{t}{b_i}\right)^{a_i}\right] \right\}.$$

The survival function is then written as

$$S(t) = \sum_{i=1}^{3} c_i \cdot \exp\left[-\left(\frac{t}{b_i}\right)^{a_i}\right].$$

The MTBF in the Weibull model is more complex than in the simple case of purely statistical failure, and is $\text{MTBF} = b \cdot \Gamma \cdot \left(1 + \frac{1}{a}\right)$.

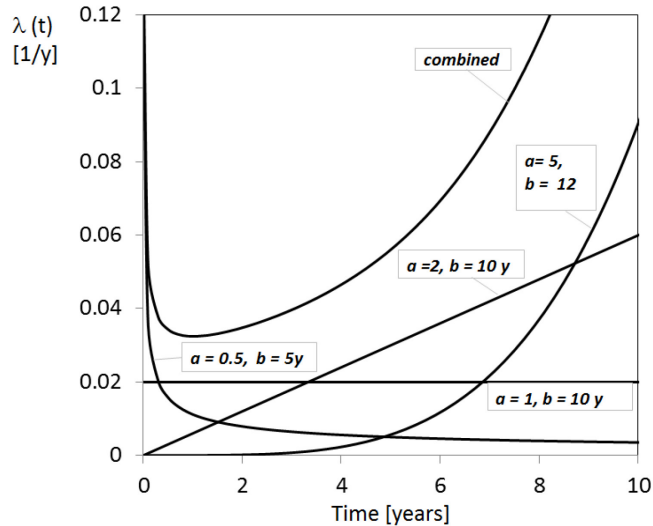

**Fig. 1:** Instantaneous failure rate in the Weibull model for different form factors $a$ and lifetime $b$

## 3 Accelerator design for high reliability

### 3.1 General remarks

In Section 2, we discussed the three modes of failure: premature, statistical, wear-out, and ageing. Combining the three modes of failure in one graph, one obtains the so-called bathtub curve (see Fig. 2) with an initially enhanced failure rate, a steady but lower failure rate in the middle and an enhanced rate at the end of the components' life cycle. The way the accelerator complex is designed, constructed, and operated has a large impact on the failure rate in all three phases.

Premature failure can be reduced by careful quality assurance and inspection of purchased components. Suppliers should be chosen based on proven reliability records and good workmanship. Often, hidden damage occurs during transport. The purchasing contract should include requirements for shock, temperature, and humidity detectors as part of packaging the components for transport. Considerable effort should be invested in acceptance testing the equipment. The tests should be comprehensive but safe, as damage of sensitive equipment during acceptance testing might occur.

The wear-out and ageing phase can be strongly influenced by regular maintenance, preventive maintenance, operating the equipment below maximum power rating, controlling the installed equipment's environment in terms of temperature, humidity, and dust, and exposure to aggressive chemicals.

All three phases of failure, however, are influenced by how the accelerator, and its subsystems and components, are designed. The following sections will discuss various aspects of high-reliability accelerator design.

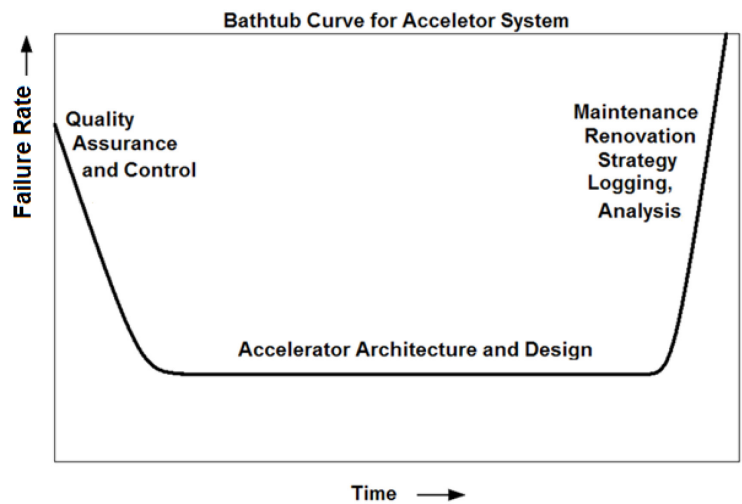

**Fig. 2:** The 'bathtub' curve of system failures over the entire life cycle

The overall optimization of an accelerator system is a compromise between three major considerations, which may lead to conflicting requirements. These are cost, performance, and reliability. The challenge of building a new accelerator facility is to find solutions that support all three requirements satisfactorily.

## 3.2 High-reliability design considerations

Next, we discuss some of the main considerations of high-reliability design. This will be followed by a more detailed discussion of examples.

### 3.2.1 *Overall complexity*

The more complex a system is, the higher the probability for hidden errors, or wrong or incompatible sets of parameters. These errors may 'sleep' within the system up to the point where special parameters in particular configurations are required. Troubleshooting in complex systems can be very time consuming. So, the complexity and interdependence of systems, which are, of course, unavoidable, should be minimized in a high-reliability design.

### 3.2.2 *Unavoidable weakness*

Some weakness in the design with respect to high reliability is thus unavoidable. For example, the accelerator beam will most probably be lost if there is a failure of a magnet power supply or if the RF system trips. The designers should be aware of these weaknesses and mitigate the risk of failure, for example, by reducing the number of individual supplies or installing spares as hot spares.

### 3.2.3 *Subsystem architecture*

The architecture of the subsystems, which defines their interdependence and their common failure modes, plays an important role in mitigating reliability risks. The system should be designed and configured such that failure in one component does not start a chain of failures in other components.

### 3.2.4 *Fail-safe design*

Provision for component failure is a part of every good design, to avoid collateral damage in case a failure occurs.

### 3.2.5 *Overrated design*

An effective way to achieve good reliability is overrated design. High-power components, such as magnet power supplies, RF transmitters, RF cavities, and pulsed magnet systems, are particularly susceptible to failure, with a potential for collateral damage. Operating these systems at the highest performance implies that operating at the temperature limit or with the maximum tolerated mechanical vibrations will shorten the components' lifetimes and increase the probability of failure. However, overrating goes along with increased cost and may not always be affordable.

### 3.2.6 *Environmental impact*

Changing environmental conditions, for example, by varying temperature, increasing humidity, changing from a dry and dusty environment to a humid environment, or exposure to aggressive chemicals (for example, created by synchrotron radiation) are important factors of failures. Protecting the equipment from such influences will be a major contribution to high reliability.

### 3.2.7 *Built-in redundancy and hot spares*

Built-in redundancy is an effective way of minimizing downtime, even if a failure does occur. Redundancy, however, may be quite costly. Hot spares are spares that need to be available anyway, to avoid long downtimes during repair or replacement but that are already installed place, ready to be used if the primary component fails. The combination of the two concepts is very effective, to avoid large downtimes and the additional cost may be relatively moderate.

### 3.2.8 *Built-in diagnostics*

Built-in diagnostics, preferably with post-mortem analysis capabilities, is a strong tool to detect potential failures before they occur or to identify quickly the root cause of failure, thereby speeding up repair and recovery from failure. Built-in diagnostics comes with a cost and must be compromised in a cost-effective design. In each case, the designs should have provisions for integrating built-in diagnostics at a later date without the necessity for major design changes.

### 3.2.9 *Repair and maintenance-friendly design*

Failures do happen in any complex technical system. This fact needs to be taken into account when designing the system. A modular concept makes it easy to diagnose and isolate the source of an error. The mechanical layout should take into account that systems need to be easily accessible for quick repair. The system must include well-documented and easily assessable measurement points for quick diagnosis.

*3.2.10 Documentation*

The effort to produce and maintain complete, up-to-date and readily available comprehensive documentation will pay for itself when errors occur in complex systems. While large documentation is usually needed for procuring and building components, the maintenance of documentation in the operational phase is often neglected, owing to lack of resources. This might be the cause of long repair and recovery.

## 3.3 Subsystem architecture

A basic choice is to choose either a compact design, which combines many functions and features implemented in the same hardware, thereby saving on redundant components, or a modular design. The modular design is friendlier for troubleshooting and repair. It also allows more easily for the incorporation of hot spares. When coupling the two types of approach, attention needs to be paid to avoid accumulating disadvantages with respect to reliable operations.

Consider, for example, a number of switched-mode power supplies. These devices are supplied with a constant d.c. voltage from individual supplies or from a common supply. The supply turns the d.c. input into a pulsed voltage with a rectangular pulse shape via a fast switch, operating at a fixed frequency in the kilohertz range. The pulse length is varied to achieve the desired output current. The output filter turns the rectangular waveform into a d.c. current.

The first, though expensive, solution is to provide a d.c. voltage supply for each individual switched-mode supply.

The second, more economical, solution is to provide a common d.c. voltage source. One can even go one step further by implementing a multichannel power supply, which has, besides the common d.c. voltage input, other common components, such as an auxiliary voltage supply for power supply electronics, and interlock and alarm features. This approach turns the switched-mode power supply system into a cost-effective compact system. The disadvantage is that most of the failures in one supply or channel will cause the common voltage supply to be shut off. All the switched-mode units supplied by the voltage source will be tripped as a consequence. Recovery involves restoring a large number of power supplies. Often, a few supplies need some extra effort to return them to service. The impact of a simple trip can be considerable. The lifetime of all the supplies turned off unnecessarily is affected.

A design developed at SLAC overcomes this disadvantage [3] by adding two isolation switches, which separate the failing switched-mode supply from its voltage source. The voltage supply thus remains turned on and all the other switched-mode supplies are decoupled. While this adds both cost and complexity, it appears to be a good compromise and will achieve considerably higher reliability performance.

## 3.4 Fail-safe design

Fail-safe design is good engineering practice, to protect a device from its own failures. However fail-safe designs that might provide optimum protection of the device might not be favourable for high reliability. Perfect protection of a device implies many trips, with many of them not being necessary to protect the device. Figure 3 illustrates the dilemma. Consider two redundant sensors for the protection of a device. Combining the two signals via a logical AND gate establishes an enhanced reliability system with little chance of false trips, but the protective function is not perfect and a faulty sensor might lead to damage of the device. This case is not fail-safe. In the opposing case, where the two signals are combined by a logical OR gate, the device is fairly well protected but false signals will lead to unnecessary trips. Reliability is compromised in this case. A system with three sensors will mitigate the shortcoming if combined as shown in Fig. 4.

Variable trip thresholds are also a good way to overcome the fail-safe dilemma. Early in operation, when there is little or no experience with the device, the trip threshold can be set low. After operational experience has been gained and the entire operation is matured, the thresholds can be raised, thereby reducing the probability of unnecessary trips. These must be part of the design from the beginning. It is also important to design a system to administer and safeguard parameters, to avoid equipment damage resulting from incorrect parameters.

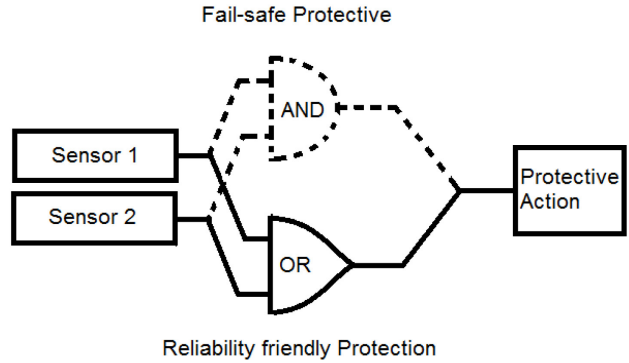

**Fig. 3:** Fail-safe versus high-reliability protection

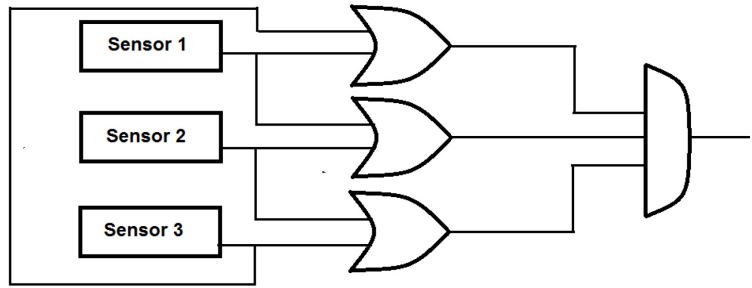

**Fig. 4:** Fail-safe and high-reliability arrangement of three sensors

## 3.5  Built-in redundancy

Built-in redundancy has the potential for improving the reliability of a system considerably. It will not reduce the failure rate but it will reduce the MTTR significantly. It might also offer a convenient way to perform preventive maintenance while keeping the system running. Closely related to built-in redundancy are hot spares. Hot spares are spare components that are installed in the system and can be switched into service without much effort and in minimum time.

Consider a system with two redundant components, labelled 1 and 2, with failure functions $F_1(t)$, $F_2(t)$. The system fails if the two components fail in the same period of time. The components will be redundant only if the failed redundant components are repaired immediately after discovery of failure. This implies that all inactive components of a redundant system need to be checked for full functionality at regular intervals. Let the time between functional tests be $t_c$. Another assumption needed for full redundancy is that failures of the redundant components would be completely independent and uncorrelated (which in a real-use case would have to be verified carefully). The redundant system will fail if the two systems fail simultaneously within a time $t_c$. The corresponding failure function $F(t_c)$ is

$$F(t_c) = F_1(t_c) \cdot F_2(t_c) = \left(1 - \exp\left[-\int_0^{t_c} \lambda_1(t')dt'\right]\right) \cdot \left(1 - \exp\left[-\int_0^{t_c} \lambda_2(t')dt'\right]\right).$$

Let us consider a well-matured system with constant (statistical) failure rate,

$$\int_0^{t_c} \lambda_1(t')\mathrm{d}t' = \lambda_1 \cdot t_c,$$

and let us assume that $t_c \ll$ MTBF or $\lambda_1 \cdot t_c \ll 1$. In this case, the failure function is approximated as
$$F(t_c) = \lambda_1 \lambda_2 t_c^2.$$

Thus by the choice of $t_c$, the failure probability of the redundant system can be made arbitrarily small under the assumption of uncorrelated failures. The MTBF$_r$ of the redundant system is then

$$\mathrm{MTBF}_r = \left(\frac{\lambda_1 \lambda_2 t_c^2}{t_c}\right)^{-1} = \frac{\mathrm{MTBF}^2}{t_c}.$$

For, example, consider a system of two redundant components with a MTBF of 480 h for each component. If the system is checked daily, the probability of failure will be $(24/480)^2 = 1/400$ and the MTBF$_r$ of the redundant system is 9600 h.

If there are $n$ parallel components that need to be active simultaneously, the MTBF$_s$ for the system is $n$ times the MTBF$_c$ of a single component, according to the considerations in Section 2. However, if there is one built-in (hot) spare, the MTBF$_s$ of the system is given by

$$\mathrm{MTBF}_s = \frac{\mathrm{MTBF}_c^2}{t_c \cdot n}.$$

Figure 5 shows as an example a crate with five switched-mode power supplies developed at DESY in 2000 (J. Eckoldt, private communication), one of which is redundant. Assuming a MTBF of 100,000 h for each supply, and assuming that the function of the redundant supply is checked once per month, the MTBF is improved to 347,000 h by using the redundant component. Thus, with an increase of less than 25% in cost, the reliability of the device is increased by a factor of 3.5.

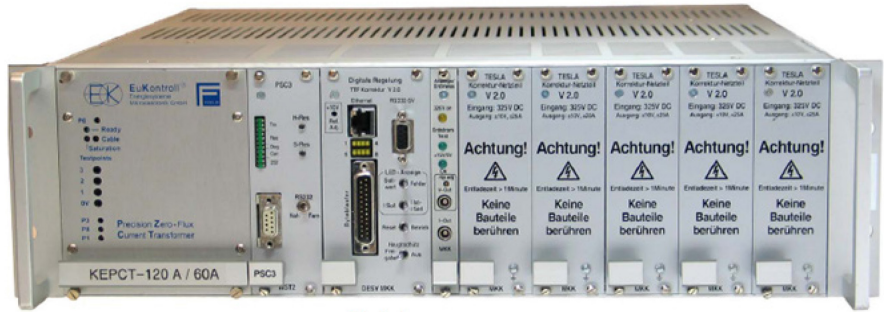

**Fig. 5:** Crate with five switched-mode power supplies, one of them being redundant (courtesy of J. Eckoldt, DESY)

### 3.6 Redundant safeguards

The case of pairs of redundant components can be easily generalized for a system of $n$ safeguards. Each individual safeguard (labelled $i$) has a mean time between failure of MTBF$_i$. Each safeguard is checked at regular intervals $\Delta t$. The MTBF$_s$ for the redundant system of safeguards is then

$$\mathrm{MTBF}_s = \frac{\prod_{i=1}^n \mathrm{MTBF}_i}{\Delta t^{n-1}}.$$

This shows that by increasing the number of safeguards, the reliability of the redundant system is increasing enormously. However, this is only true if $\Delta t <$ MTBF$_i$, that is, if the components are checked at regular, sufficiently small, intervals. It is also important that the redundant components are as diverse as possible to reduce the probability of correlated failures.

## 3.7 Overrated design

Overrating of high-power components has a number of positive effects on the lifetime and probability of failure of the components. These have mainly to do with operating temperature. The operating temperature will be less in an overrated device as the cooling is designed for larger power dissipation than the power dissipated in normal operating conditions. The change in temperature when the device is turned on and off will also be reduced. This leads to reduced mechanical stress. All these factors will increase lifetime and reduce failure rate. In the electronics industry, the following Arrhenius-based expression is used to describe the impact of temperature and temperature changes on the failure rate [4]:

$$\frac{\lambda}{\lambda_0} = \left(\frac{\Delta T}{\Delta T_0}\right)^2 \cdot \exp\left[-\frac{E}{kT} \cdot \left(1 - \frac{T}{T_0}\right)\right].$$

The first factor describes the effect of thermal cycling and fatigue; the second factor describes the thermal stress due to high temperature. The index '0' indicates a reference temperature level for comparison. The parameter $k$ is Boltzmann's constant, and $E$ is a typical excitation energy level that leads to changes in the material under consideration

Thus, overrating leads to a reduction in the failure probability. Another positive effect is that the system can be operated further away from critical limits and trip thresholds, thereby reducing the number of false trips. On the negative site is increased cost and, in many cases, increased installation space requirements.

## 3.8 Environmental impact: dust, humidity, temperature

In Section 3.7, we discussed the impact of temperature and temperature changes on the lifetime and on the probability of failure. The impact of temperature on the lifetime of electrolytic temperature is pronounced and well understood. Manufacturers quote the following expression for the lifetime, based on Arrhenius law (see for example, Ref. [5]):

$$\text{MTBF}(T) = \text{MTBF}(T_\text{ref}) \cdot 2^{-\left(\frac{T - T_\text{ref}}{10\,°C}\right)}.$$

Figure 6 shows how the lifetime of film capacitors is affected by internal temperature, which is determined by ambient temperature and internal heat production. Similar behaviour is observed for electrolytic capacitors.

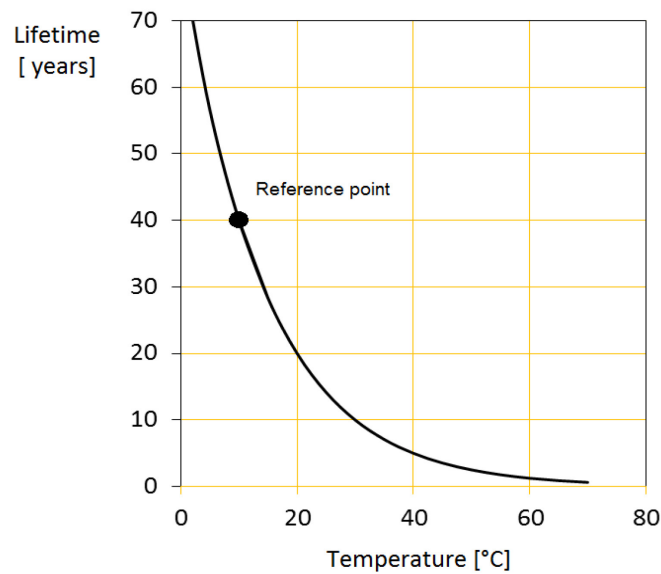

**Fig. 6:** Lifetime of film capacitors as a function of internal temperature

Exposure to changing humidity and dust are other environmental factors that can significantly compromise the reliability of technical components.

Dust may have constituents that are electrically conducting and could cause electrical shorts between connectors on electrical printed circuit boards. Once a current is flowing, the printed circuit board material can become carbonized, which aggravates the short to the point that the electronic component becomes non-functional. Needless to say, such a process can be accelerated by the presence of humidity or of chemical radicals in the air. Such radicals may be produced by synchrotron radiation, which is produced by high-energy electron beams in the accelerator tunnel. The malfunction in such cases may develop only gradually and slowly and may initially only cause occasional failures. The cause of such failures can be very difficult to find and repeated failures may cause significant downtime. Figure 7 shows a board in the quench protection system of the Tevatron (Fermilab) superconducting magnet system that was damaged by this effect (H. Edwards and P. Czarapata, Groemitz Miniworkshop on Accelerator Reliability (Groemitz) 2005, unpublished data).The combination of dust and humidity caused some mysterious modulator trips in the RF system of the HERA electron–positron collider at DESY, Hamburg. In the late 1990s, a high frequency of arcing on the modulator was reported. In an attempt to explain the events, they were analysed as a function of air humidity and dust particles in the air. There was no clear correlation. However, after further investigation, it was found that the arcs always occurred when the outside weather conditions changed from dry to humid periods. A small amount of outside air had been added to the circulating, well-conditioned, internal airflow. During dry periods, small amounts of dust from the outside air accumulated on the surface of high voltage (≈70 kV) carrying components. When the humidity increased, the dust particles acted as launch points for arcs into the slightly more humid air in the modulator room. Thus, even a tiny amount of exposure to environmental conditions can have surprisingly large effects.

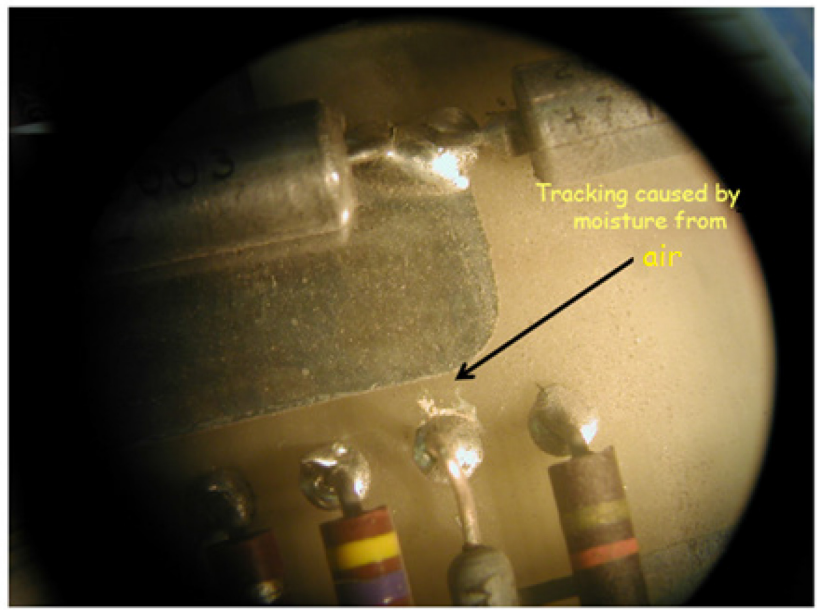

**Fig. 7:** Damage to quench protection board as a result of dust and humidity

An effective, though not quite inexpensive method, was chosen to mitigate the environmental impact on electronics in the new synchrotron light source, NSLS-II, at Brookhaven National Laboratory. All electronics components are enclosed in a sealed rack system, in which air is circulated around the equipment for cooling and is then cooled with water-to-air heat exchangers. This method has the advantage of keeping cooling water away from the magnet power supply and other electrical equipment and maintaining air cleanliness and low humidity. As power dissipation is low in all devices, air cooling is sufficient to maintain favourable operating temperatures. Figure 8 shows the NSLS-II rack system.

## 3.9 Error-prone solutions

In reviewing the cause of failure in accelerator equipment, there are two outstanding items. The first one is leaking cooling water, which can cause electrical shorts, damage of electrical equipment, or the formation of acidic liquid, which will lead to corrosion. While large leaks are detected easily and the recovery from a large leak is straightforward, small leaks may remain unnoticed and a large amount of damage might result before the problem is noticed. For this reason, water cooling systems have to be designed and manufactured with great care. Wherever air cooling can be used instead of water cooling, air cooling should be given preference. Water piping should, if possible, be installed underneath delicate equipment rather than overhead. Double floors that provide a space to accommodate water piping are an expensive but effective solution to avoid damage by cooling-water accidents. An example of a water-free cooling design is the air cooled sealed NSLS-II power supply enclosure mentioned previously (see Fig. 8), which keeps cooling water away from electrical components. Water cooling circuits should be regularly checked for pressure drop and the design should foresee an efficient way of performing such tests frequently.

Another frequent source of failure is cable terminations and connectors. The list of potential issues is long and spans from assembly errors, insufficient ground connections, corrosion, and mechanical damage, cable damage at the fitting, miswiring, and confusion of connectors after repair or test. The root cause is that cable connections often have to be mounted in the field under sometimes difficult conditions, such as limited space, visibility, accessibility, dust, etc. For this reason, it should be checked from case to case whether analogue hard-wiring can be replaced by digital data connections, which offer the ability to replace many critical electromechanical connectors and switches. To save cost on cable connections by using cheap components and inexpensive labour might not be an optimum decision in view of the loss in reliability caused by low-quality connectors and poor workmanship. It is advisable to design a comprehensive quality-control programme to assure the adequacy of cable connections. To reduce the risk of slowly developing poor connectivity, the use of gold-plated connectors is encouraged. In the case of high-current bolted cable connectors, the materials have to be carefully chosen to match in thermal expansivity, to avoid premature wear-out of the bolting elements.

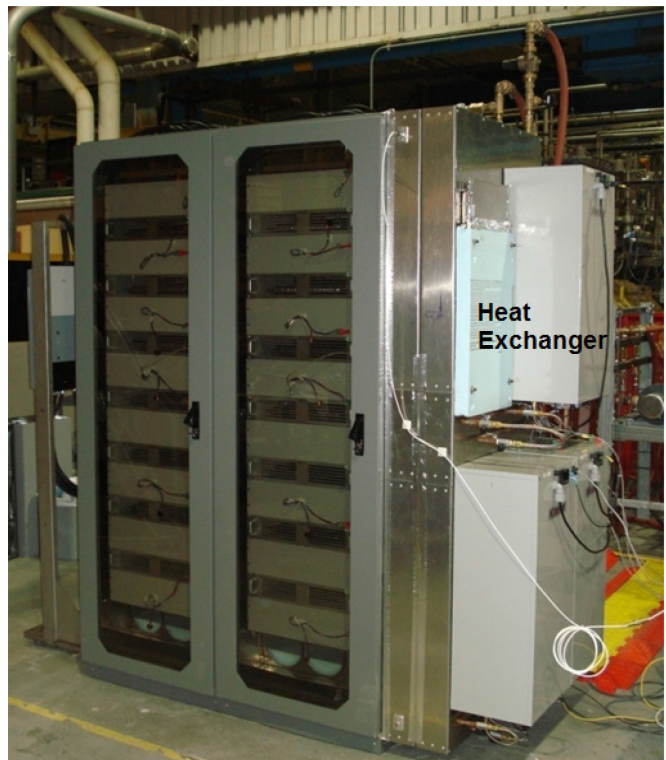

**Fig. 8:** NSLS-II sealed racks (equipment enclosures)

## 3.10 Built-in diagnostics

The next two topics are strongly related to operation and maintenance of high reliability, but have to be considered during the design and construction phase of a facility. The first topic is built-in diagnostics. Usually, a failure announces itself by more or less significant changes in some of the analogue variables inside a device. Therefore, it is helpful in both preventing and analysing a failure if the internal analogue variables of a technical device are measured, recorded, logged, distributed, and analysed. Thus, built-in diagnostic tools have the potential to increase the MTBF and reduce the MTTR of a component or system.

Comprehensive capturing of relevant internal data has to be well integrated in the design of a device, component, or system. With increasing complexity of a technical system, the need of internal diagnostics for supporting preventive maintenance and troubleshooting becomes more crucial. There are many examples of good implementation of built-in diagnostics. The cost of the additional design and construction effort is not negligible. However, it will be extremely difficult to achieve reliabilities above 95% without built-in diagnostics. For these reasons, built-in diagnostics is a standard component of modern technical designs.

## 3.11 Repair- and maintenance-friendly design

Accessibility for troubleshooting and repair is an important aspect of service-friendly design in support of high operational reliability by quick recovery from failure. Another service-friendly design feature is modularity. Modularity refers to the ability of independent testing for proper functioning of a part of a system without major rearrangement, partial disassembly, or reconfiguration. Modularity also refers to the physical arrangement of the constituents in a way that allows a larger subsystem to be exchanged with a minimum of effort. A thoughtful design foresees easy access to internal components for taking measurements during troubleshooting. This includes the provision of circuit board extensions to easily access measurement points, or the provision of a connector that can be connected to a test module for automated troubleshooting.

## 4 High-reliability operations

## 4.1 General remarks

Reliable operations that are based on a high-reliability design and implementation are achieved by a process of continuous improvements. Many subsystems of an accelerator facility are custom-designed systems with few components. They will mature during operation by small improvements and partial replacement of weak components. While this process is expected to require a significant effort in the start-up years of a facility, it will settle to a lower level of effort after a few years but will persist up to the time when larger investments and refurbishment become necessary, when the components arrive at the end of their life cycle.

An important tool for organizing efficient and reliable operation is comprehensive data logging and analysis of the performance of all components. The analysis toolkit will usually not be provided as part of the construction but its creation is an iterative process, which also requires a certain amount of operation time to reach a sufficiently high level of maturity.

The logged data need to have timing information and the systems should be equipped with circular buffers to allow post-mortem analysis. Data should be easily accessible from off-site to allow experts to perform analysis and troubleshooting without their having to be physically on-site.

Root cause analysis is helpful in understanding larger incidents, to prevent them from happening again. Commercial software tools are available to support such activities.

High operational reliability requires well thought-through operational strategies to mitigate the impact of failure, in particular, in view of the always-limited operational resources. Important elements of such strategies are scheduled maintenance and a preventive maintenance programme based on comprehensive monitoring of components and analysis of the data, as discussed previously.

One way of optimizing the facility output is to develop a figure of merit for operational performance and use this to relate any component failure to reductions in performance. This enables one to decide rationally whether to interrupt operations for an unscheduled intervention or to run with reduced performance until the next scheduled intervention. Part of such strategy is the inclusion of back-up plans for operating with reduced performance, such as accelerator studies or special operation modes.

### 4.2 Preventive maintenance

With the large number of components and the large diversity of equipment in a large accelerator facility, the opportunity for preventive maintenance to replace, repair, or adjust equipment before failure occurs is large. Systematic preventive maintenance of each piece of equipment is, in most cases, unrealistically expensive, and the effectiveness of such activity is very poor. Preventive maintenance, therefore, needs to be properly focused on equipment and use-cases where it will have a high probability of being effective in preventing failure. In this section, we will discuss preventive maintenance opportunities that have proved effective.

Mechanical rotating equipment is an obvious candidate for preventive maintenance. Examples of such equipment are water pumps, mechanical vacuum pumps, fan systems, air compressors, chilled water compressor systems, and turbines in cryogenic cold-engines. Such equipment is needed in many industrial installations. Such systems usually come with detailed maintenance plans from the manufacturer. Often, preventive maintenance is offered as part of a service package by the manufacturer. The preventive maintenance of such systems has been standardized and standards published, for example the Society of Automotive Engineers JA1011 standard *Evaluation Criteria for Reliability-Centered Maintenance* [6].

Another obvious class of equipment are battery-based devices, such as uninterruptable power supplies. Batteries have a well-known lifetime in terms of operating hours. Thus, preventive maintenance is very effective in such cases.

It is also obvious that filter systems need cleaning or filter replacement at regular intervals, which depend somewhat on the environment in which they have to work. Maintenance plans can be optimized after short operation periods.

Much equipment is cooled by air driven by internal fan systems. Fans should be interlocked so that in case of failure, there is no consequent damage to high-power equipment. The lifetime of fans should be known, in principle. However, the range of achieved operating hours is quite large. The performance and lifetime of fans depends on humidity, dust, and ambient temperature. Therefore, some judgement and experience is required to know when to exchange fans preventively, to avoid excessive costs or to prevent shutdown of equipment during operations.

The preventive maintenance that can be carried out on electrical equipment, such as magnet power supplies, is somewhat less obvious. Often, these systems have clamps or bolds to hold equipment in place. These mechanical fixtures need to be checked from time to time to ensure that all equipment is tightly connected, to avoid arcing and other damage to high-power equipment. An efficient method to check the integrity of connections that carry a high current is to use thermal imaging (see Fig. 9), which allows a large amount of equipment to be monitored in a very short time. Critical equipment can also be monitored continuously by a fixed installation; the cost of thermal imaging has reduced dramatically in the recent decade.

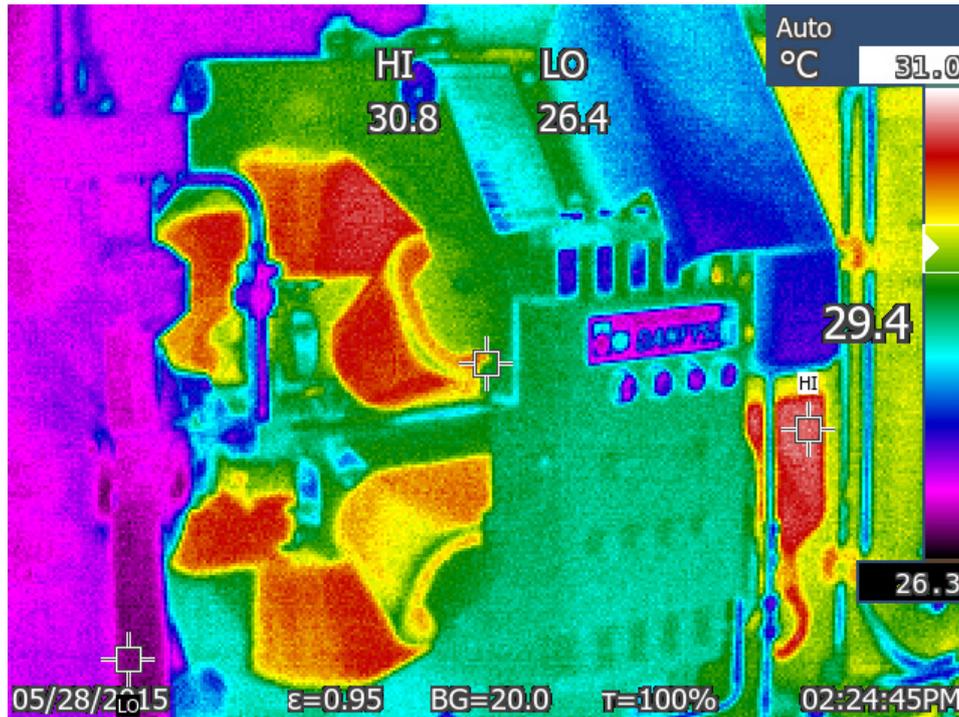

**Fig. 9:** Screenshot of thermal imaging of high-current connectors and magnet coils which is a very efficient method to check for weak connectors, obstructed cooling channels on magnet coils, etc.

Water cooling piping at or inside equipment should be regularly checked, since, as pointed out, cooling-water leaks are often the root cause of malfunction. A static pressure test will easily detect cooling-water leaks that can cause significant damage or downtime.

Some equipment exhibits sign of wear-out and fatigue. Some thyratron tubes in pulsed power equipment announce the end of their lifetime by requiring higher voltages to maintain the discharge current. Poor contacts also show increased transition resistance. Latent winding faults on magnet coils can be recognized by carefully examining inductive voltages and comparing them with electrical current changes.

The lifetime of piping systems and cooling channels inside equipment is very difficult to estimate. Here, it is important to check the water quality—its resistance, oxygen content, and pH—on a regular basis. It is also very difficult to estimate the impact of synchrotron radiation on the cooling water and the cooling channels. Variable thermal loads may lead to repeated thermal stress, which can cause fatigue and subsequent cooling-water leaks, which can be detected by static pressure testing on isolated cooling channels.

Maintenance is labour-intensive and is one of the highest cost elements in operating an accelerator. It is important that precious resources are used in the most effective way. This requires that maintenance programmes to focus on components with a high-failure probability. Error and failure analysis, supported by modelling, can be helpful tools to develop effective maintenance programmes.

During the start-up of operations of a new facility, it is usual for a large number of teething issues on new systems to be addressed; it is difficult to introduce a systematic preventive maintenance programme at the same time, owing to limited resources and a lack of data on failure events. Preventive maintenance is thus most effective in the mature operation phase. Preventive refurbishment is naturally most reasonable if the systems enter the wear-out phase. However the time constants for the subsystems are probably different, so that not all operational reliability phases occur simultaneously.

Next, we will demonstrate, with an example, how the formalism of reliability engineering can be applied to decision-making in maintenance activities. Consider a system of 200 components that are subject to wear-out. Figure 10 shows the available data on failures that have occurred in components within a few years. The failures start to appear after about 200 weeks and the failure rate afterwards is accelerated. Do these observations suggest that preventive refurbishment will avoid considerable downtime in the future? The data are well described by a two-parameter Weibull failure function

$$F(t) = 1 - \exp(-(t/\tau)^\alpha),$$

the parameters $\alpha$ and $\tau$ being obtained by a least square fit. We then calculate the mean residual life:

$$\mathrm{MRL} = \int_0^\infty dt \cdot \frac{S(t+t_0)}{S(t_0)}.$$

This allows us to project future failure rates and enables us to make a rational decision on preventive refurbishment, to prevent unscheduled downtime (see Fig. 11). The MTBF is about 494 weeks and the form factor is $\alpha = 6.1$. Thus, the system is described well by failures in the wear-out phase. The mean residual life at the end of the observation period is reduced to only one-third of the original system and is expected to become very small after another operation time of ~200 weeks, which constitutes a high risk for unscheduled downtime. In this situation, preventive refurbishment might be considered an overall optimum measure.

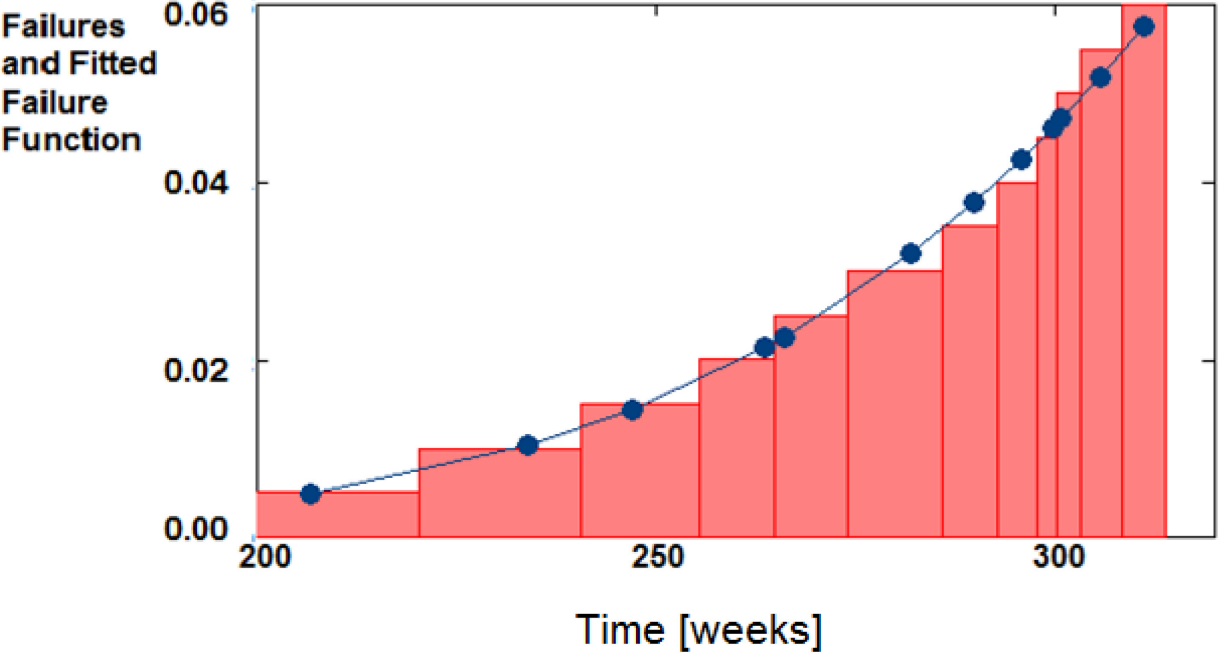

**Fig. 10:** Example of accumulated failures (columns) and Weibull Failure function (two-parameter fit, dots)

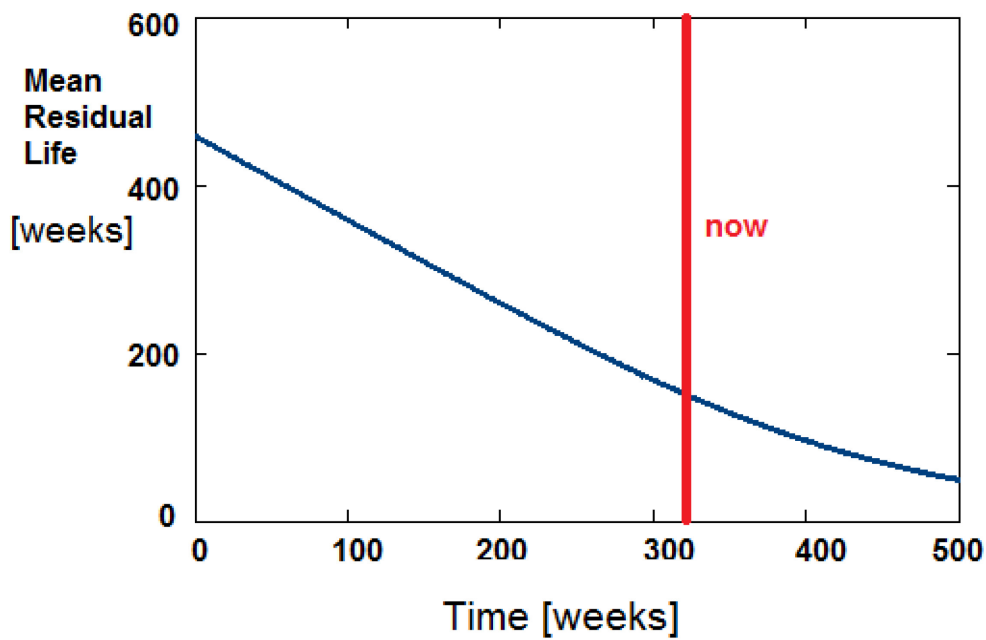

**Fig. 11:** Mean residual life as a function of operating time. The mean residual life is expected to drop significantly after additional operation time.

### 4.3 Speed-up of repair

The MTTR includes the time to identify a malfunctioning component. This can be quite time consuming and cumbersome, as trips and faults that are observed are often only the consequence of a hidden primary malfunction. This time may exceed the time for repairing or exchanging a component. The information provided about a failure event is thus an important factor in the overall system reliability. The following measures have proven to be very helpful in speeding up the troubleshooting and investigation process:

— **Transient recording** is based on continuous monitoring with an appropriate data rate, which may have arbitrarily short time-scales. Data storage is limited, and this defines a data cycle in which the oldest data are continuously overwritten by new data (also called a 'circular buffer'). The failure event needs to provide a trigger that stops the circular buffer and data for the time period that precedes the event that can be retrieved (post-mortem data).

— An **asset management** database with life cycle data on each component may be very helpful in finding the root cause of a failure event. Knowledge of a component's history (previous failures, unusual operating conditions, time in operation, problems with similar equipment) provides, in some cases, a clue to understanding a malfunction.

— **Remote access** to the process data generated by a system or an item of equipment is important, as it tends to save time and cost, since experts may then not need to come on-site to perform troubleshooting or to direct repairs, so long as they have access to the Internet somewhere on the planet.

— Having the data from a failure available is sometimes not sufficient. **Analysis tools** are likely to be required if a large amount of data needs to be scanned for anomaly behaviour of the equipment.

— It is worthwhile to maintain a **database with information on failures**, so as not to be solely dependent on the memory of experts involved in resolving past problems.

— **Start-up checklists** are helpful, to find malfunction early and to avoid having to repeat a start-up procedure due to late discovery.

## 4.4 Spares inventory

The availability of spares is crucial for high-reliability operations, as many spare parts require a long time for replacement or repair. Therefore, for all breakable equipment of an accelerator facility, there should be at last one spare, which is best procured as part of the construction project. For equipment with a purchasing lead time of the order of the system MTBF, (which is $n$ times smaller than the single component MTBF, $n$ being the number of pieces of equipment in the system) more than one spare is necessary to avoid extended downtimes.

To plan highly reliable operations, it is necessary to determine the rate of consumption of spares. The replacement of used spares is a major cost factor in an operations budget. For a system with identical components, the failure density distribution of which may be described by the Weibull model, the probability of failure of each component in the interval [0,$t$] is described by

$$F(t) = 1 - \exp\left[-\left(\frac{t}{b}\right)^a\right]$$

and the probability of surviving longer than $t$ is $S(\tau) = 1 - F(t)$. The lifetime $\tau$ of a component is defined as the time when $S(\tau) = 1/e$. It is identical to parameter $b$ in the distribution:

$$1 - \frac{1}{e} = 1 - \exp\left[-\left(\frac{\tau}{b}\right)^a\right] \rightarrow 1 = \left(\frac{\tau}{b}\right)^a \rightarrow \tau = b.$$

Where $a = 1$ (statistical failures), $\tau$ is identical to the MTBF and the lifetime of a system of $n$ identical components is $\tau/n$, as shown in Section 2. The annual replacement of spares of a system of $n$ identical components is then $n/(\tau/1 \text{ year})$.

However, this consideration is only correct where there is an equilibrium in decay and replacement. If we have a component with a long lifetime, let us say 20 years, the probability for failure in the first few years is very small. Thus the yearly reinvestment of cost per unit $\times$ $n/\tau$ is unnecessary.

Therefore, for adequate but economical repurchases of spares, we need to take into account that we are starting with a complete spare inventory and that it might take years before the available spare is used and needs to be replaced. A compact formula for the replacement of spares over time is given by Hoffstaetter and Willeke (unpublished data) and is briefly presented in the following.

Let us consider a system with $n$ identical components. The failure of components of the system is described by the function $F(t)$, which may be modelled using a Weibull distribution. This assumption is not necessary for the assessment we are going to make, but it will allow us to calculate examples easily, to emphasize the importance of the considerations that will follow.

We remember that the failure density distribution function $f(t)$ is the time derivative of $F(t)$. We will further define the rate of component replacement as $R(t)$

$$f(t) = \frac{\mathrm{d}}{\mathrm{d}t} F(t).$$

We now consider replacements of failed components. The number of initially installed components to be replaced after some step in time $\delta t$ is $R_0(t)\cdot\delta t = f(t)\cdot\delta t$. However, the replaced parts $f(t')\cdot\delta t$ with $0 < t' < t$ are also subject to failure, at a rate $f(t - t')$ at time $t$, and we have to add these to the list of replacements (the infinitesimally small factor $\delta t$ will be dropped from now on):

$$R_1(t) = f(t) + \int_0^t f(t') \cdot f(t - t')\mathrm{d}t'.$$

The replacements of the already replaced parts, however, are also subject to failure, so we are left with an infinite number of nested integrals:

$$R(t) = f(t) + \int_0^t dt' f(t') \left[ f(t-t') + \int_0^{t-t'} dt'' f(t'') \left[ f(t-t'') \int_0^{t-t'-t''} dt''' f(t''') f(t-t''') \ldots \right] \right]$$

The replacement at time $t$ replaces initially installed components with failure rate $f(t)$, and components that were replaced at any earlier time $t'$ are replaced at the rate $f(t-t')$. This consideration leads to the compact integral equation

$$R(t) = f(t) + \int_0^t dt' R(t') \cdot f(t-t').$$

Note that the function $f(t)$ can be assumed to be zero for $t < 0$. Any other value does not make sense, since $f(t)$ describes the failures of parts that did not yet exist at $t < 0$. Parts that do not yet exist also do not have to be replaced, so that $R(t<0) = 0$ as well. For this reason, the integrand is
$$R(t') \cdot f(t-t') = 0 \text{ for } t' < 0 \text{ and for } t' > t.$$

Thus, we can rewrite the equation as
$$R(t) = f(t) + \int_{-\infty}^{\infty} dt' R(t') \cdot f(t-t').$$

We use the fact that the Fourier transform of a convolution
$$\int_{-\infty}^{\infty} dt' f(t') \cdot R(t-t')$$

in the time domain equals the product of the Fourier transform of $f(t)$ and $R(t)$. With

$$(\tilde{f}(\omega), \tilde{R}(\omega)) = \frac{1}{\sqrt{2\pi}} \int_{-\infty}^{\infty} (f(t), R(t)) \cdot \exp[-i\omega t] dt.$$

We arrive at
$$\tilde{R}(\omega) = \tilde{f}(\omega) + \sqrt{2\pi} \cdot \tilde{f}(\omega) \tilde{R}(\omega),$$

with the solution
$$\tilde{R}(\omega) = \frac{\tilde{f}(\omega)}{1 - \sqrt{2\pi} \cdot \tilde{f}(\omega)}.$$

The replacement function is thus

$$R(t) = \frac{1}{\sqrt{2\pi}} \cdot \int_{-\infty}^{\infty} \tilde{R}(\omega) \cdot \exp[i\omega t] \, d\omega = \frac{1}{\sqrt{2\pi}} \cdot \int_{-\infty}^{\infty} \frac{\tilde{f}(\omega)}{1 - \sqrt{2\pi} \cdot \tilde{f}(\omega)} \cdot \exp[i\omega t] d\omega.$$

Note that
$$\tilde{f}(0) = \frac{1}{2\pi} \int_{-\infty}^{\infty} f(t) dt = \frac{1}{2\pi},$$

and the integrand in the expression has a pole at $\omega = 0$, since

$$\sqrt{2\pi} \cdot \tilde{f}(\omega)_{(\omega=0)} = \int_0^{\infty} f(t) dt = F(\infty) = 1,$$

which requires some care in the integration. This pole reflects the fact that for times much longer than the lifetime $\tau$, the replacement function approaches the constant $1/\tau$, so that the time integral of $R$, $\tilde{R}(0) = \int_{-\infty}^{\infty} R(t')dt'$, tends to infinity.

The replacement function for a system with Weibull parameter $a = 1$ is the constant $1/t$ over the entire range from zero to infinity. For $a > 1$, the replacement function is zero for $t = 0$, around $t = \tau$, the rate is larger than $1/\tau$ and $t$ will oscillate around $1/\tau$ with a frequency $1/\tau$ and decreasing amplitude. For $a < 1$, the replacement function departs quickly from its start value and approaches zero very slowly. So in this case, the replacement will not approach the limit of $1/\tau$. This is due to the fact that for long times, the failure rate is approaching zero. Figure 12 shows the replacement as a function of time for various values of $a$ ($a = 3, 2, 1$, and $0.6$).

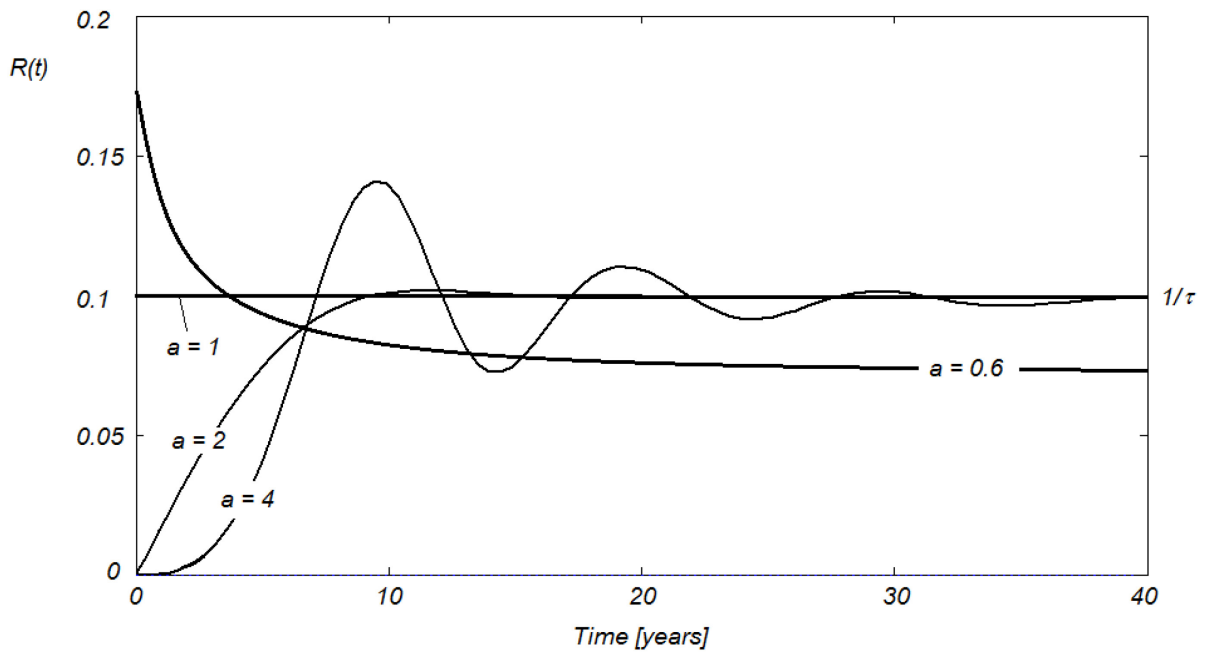

**Fig. 12:** Replacement of used spares as a function of time for various system parameters (Weibull parameter $a$) for $a = 1$ (constant failure rate), $a = 2, 3$ (failure due to ageing and wear-out), $a = 0.5$ (early failure).

Let us consider an example:

Consider a system that has a certain number of components $N$. This number $N$ is assumed sufficiently large that statistical fluctuations are smaller than the systematic trends. A fraction $c_1 = 10\%$[1] fail prematurely, with a lifetime parameter of $\tau_1 = 1$ year and a form factor $a = 0.4$. A fraction $c_2 = 30\%$ has statistical failure characteristics with a lifetime of 20 years and the reminder of the components fail because of wear-out, with a lifetime parameter of 20 years as well and a form factor of $a = 4$.

The corresponding hazard function $\lambda(t)$ and the failure probability distribution function (called here $f(t)$) are shown in Fig. 12; the replacement function $R(t)$ is shown in Fig. 13. We see that the estimated expenses per unit of time for spare replacement start out much lower than the average $1/\tau$.

---

[1] This value is intentionally chosen unrealistically large to make the effect more clearly visible in the results.

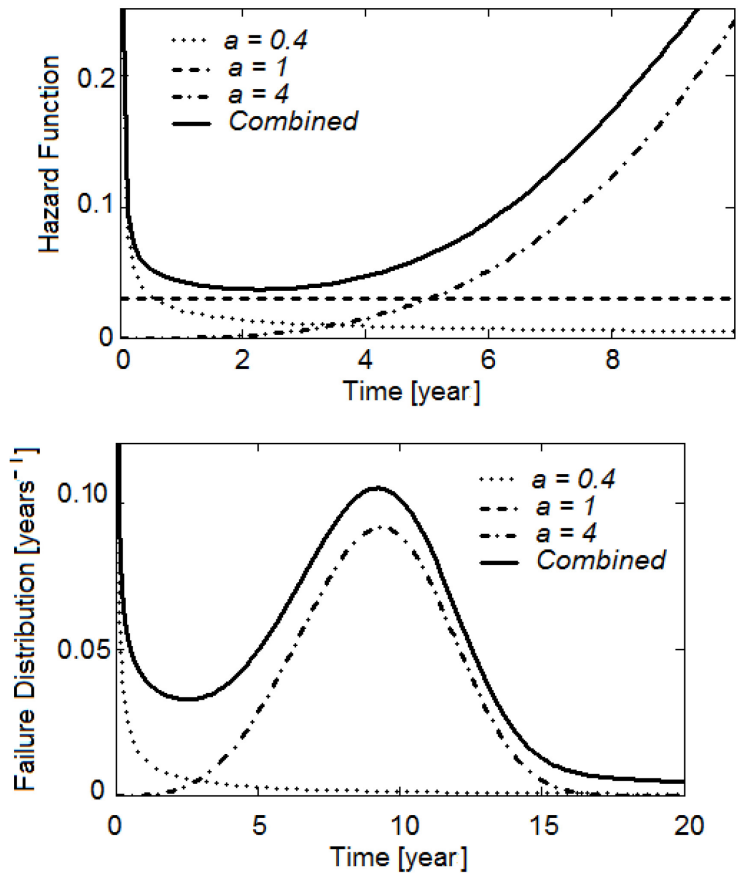

**Fig. 13:** Example hazard functions and failure probability distribution. The combined hazard function is calculated by $\Lambda(t) = f(t)/S(t)$ with $f(t)$ and $S(t)$ being the sum of the corresponding partial functions.

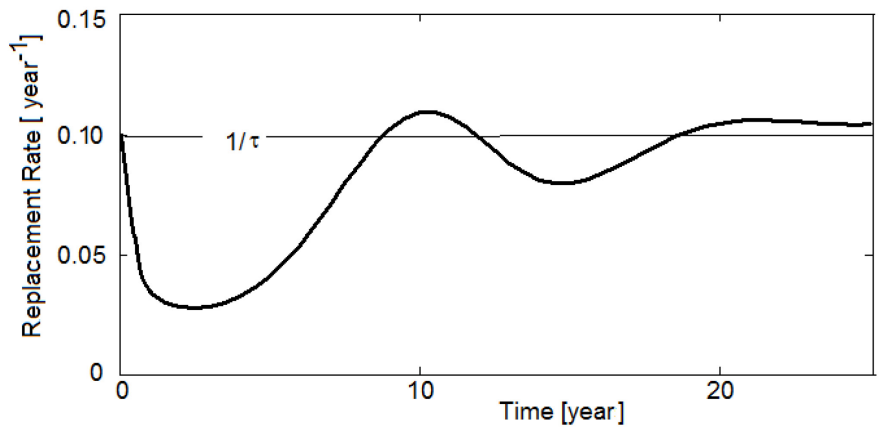

**Fig. 14:** Replacement of spares over time for a system characterized by the hazard function in Figure 13

The expected expenditure rate on spares in the first few years of operation (up to about half the lifetime) is significantly lower than the average spare consumption rate of $1/\tau$ which is obtained asymptotically for a well matured system

$$\lim_{t \to \infty} R(t) = \frac{1}{\tau}.$$

The consumption rate will exceed average spare consumption once the equipment reaches the end of its life; after this, the cost will approach the average cost while oscillating around the average.

The integrated spare replacement $I_s$ of a large period of t $\simeq \tau$,

$$I_s(t) = \int_0^t R(t')dt',$$

will remain significantly below the integral $t/\tau$ which is based on the asymptotic value of the spare consumption rate $1/\tau$ (see Fig. 15).

Thus taking the more advanced spare cost analysis into account will release substantial funds for planning and addressing other reliability issues as they occur in the early phase of a facility's life.

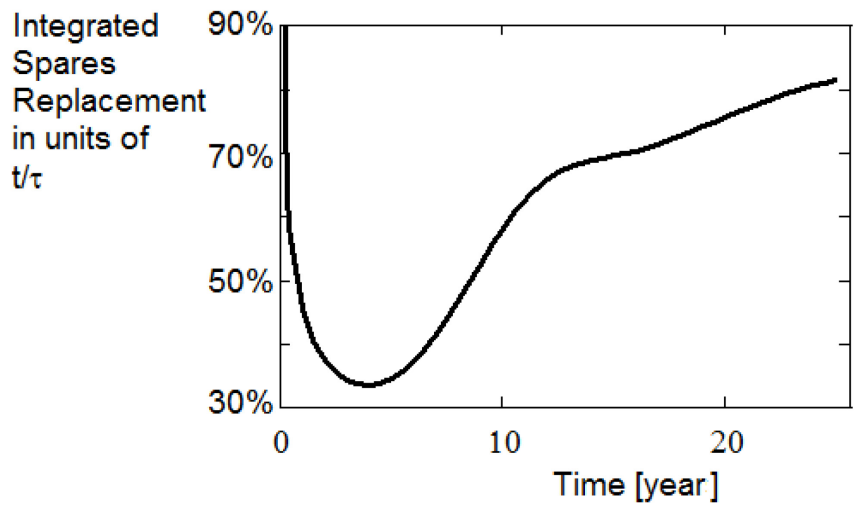

**Fig. 15:** Integrated spare replacement over time is considerably lower than the average spare replacement for a mature system with a replacement rate of $1/\tau$.

### 4.5 Human performance factors

Human performance issues are a topic that greatly exceeds the scope of this lecture. While it should not be completely omitted, it is impossible do full justice to this topic, given its importance. Here, we list the most important aspects of the impact of human error on the reliability of a facility. A facility with a nearly perfect hardware system and nearly perfect controls is fairly insensitive to human error. However, the implementation of nearly perfect systems is most certainly not the most cost-efficient implementation of operations. For this reason, human intervention is required to keep systems running and to handle exceptional events. These interventions are prone to human error.

Human errors are difficult to eliminate completely but there are measures that can be reasonably taken to minimize them.

— A clear definition of the line of command is mandatory, to handle exceptional operations. These need to be communicated clearly and all those involved in operations need to understand and accept these definitions. The line of command may not be static but may change for different phases and modes of operation. Particular care has to be taken to describe the transition from one mode to the other. In particular, the return to normal service needs to be addressed.

— The operational roles, responsibilities, and accountabilities also need to be defined, spelt out, and communicated clearly.

— Operational information needs to be distributed regularly in briefings, and shift turnover meetings, to avoid information gaps or misunderstandings, which may lead to failure and operational inefficiencies.

- It is important to set all procedure and operational rules in writing. Care should be taken to prepare such write-ups and the consensus of all those involved should be secured.
- Automation of operational procedures is advisable wherever feasible, affordable, and safe, to avoid overloading operators, especially in emergency situations.
- Facility operators should have a solid base training, which should emphasize the handling of exceptional situations.
- Comprehensive system information should be available online or in the form of hard copies.
- Operational software should take ergonomic considerations into account. The use of colour to convey information is not recognizable by everybody and should be minimized.
- Obviously a comprehensive, well optimized alarm system is essential, to recover quickly from faults and trips.
- Access to equipment and controls should be carefully optimized. While access limitations might stand in the way of quick recovery from a failure, the absence of access limitations can be a cause of failures through false settings and misunderstandings.
- Ambiguous naming is a frequent source of misunderstanding, uncoordinated action, and loss of operational efficiency. Names and labels need to be unambiguous. The information on naming needs to be communicated well; the wrong use of names may not be ignored or considered insignificant.

## 5 Closing remarks

This report intends to provide an overview of considerations that are related to high-reliability operations of accelerator-based science user facilities. It would have been beyond the scope of this report to cover each of the topics comprehensively. For this reason, this write-up should be considered as an encouragement to study the areas discussed in this report in more detail. Quite a number of textbooks and scientific and technical publications on reliability engineering, human performance issues, and system maintenance are available for further study.